# Status and direction of atom probe analysis of frozen liquids

Patrick Stender[1+], Baptiste Gault[2,3], Tim M. Schwarz[1], Eric V. Woods[2], Se-Ho Kim[2], Jonas Ott[1], Leigh T. Stephenson[2], Guido Schmitz[1], Christoph Freysoldt[2], Johannes Kästner[4], Ayman A. El-Zoka[1+]

[1] Institute of Materials Science, Chair of Materials Physics, University of Stuttgart, Heisenbergstrasse 3, 70569, Stuttgart, Germany

[2] Max-Planck-Institut für Eisenforschung, Düsseldorf, Germany.

[3] Department of Materials, Royal School of Mines, Imperial College London, London, UK

[4] Institute for Theoretical Chemistry, University of Stuttgart, Pfaffenwaldring 55, 70569 Stuttgart, Germany

[+] Corr. Authors: Ayman El-Zoka a.elzoka@mpie.de; Patrick Stender patrick.stender@mp.imw.uni-stuttgart.de

**Abstract**

Imaging of liquids and cryogenic biological materials by electron microscopy has been recently enabled by innovative approaches for specimen preparation and the fast development of optimised instruments for cryo-enabled electron microscopy (cryo-EM). Yet, Cryo-EM typically lacks advanced analytical capabilities, in particular for light elements. With the development of protocols for frozen wet specimen preparation, atom probe tomography (APT) could advantageously complement insights gained by cryo-EM. Here, we report on different approaches that have been recently proposed to enable the analysis of relatively large volumes of frozen liquids from either a flat substrate or the fractured surface of a wire. Both allowed for analysing water ice layers which are several microns thick consisting of pure water, pure heavy-water and aqueous solutions. We discuss the merits of both approaches, and prospects for further developments in this area. Preliminary results raise numerous questions, in part concerning the physics underpinning field evaporation. We discuss these aspects and lay out some of the challenges regarding the APT analysis of frozen liquids.

# 1 Introduction

Atom probe tomography (APT) provides 3D elemental mapping, typically with sub-nanometre resolution (De Geuser & Gault, 2020), and an elemental sensitivity down to the range of tens of parts-per-million (Haley et al., 2020). Deploying these capabilities of APT to study wet chemical systems has been hindered by the lack of available specimen preparation strategies for frozen liquids. Early efforts in this direction were reported (Stintz & Panitz, 1991, 1992; Pinkerton et al., 1999), however, no workflows have established the use of the APT technique for routine analyses of frozen liquids. This contrasts with transmission-electron microscopy (TEM), which over the past decades has seen tremendous developments in standardized workflows allowing preparation and handling of specimens at cryogenic temperature (Marko et al., 2007; PARMENTER & NIZAMUDEEN, 2020; Livesey et al., 1991) as well as the analysis of liquids via graphene encapsulation (Park et al., 2015), recently achieving atomic resolution (Nakane et al., 2020).

For routine APT analysis, all data is acquired at a cryogenic base temperature (typically 20-80K), but the preparation of specimens by electrochemical polishing (Miller, 2000) or focused-ion beam (FIB) (Prosa & Larson, 2017) is typically performed at room temperature. There have been efforts to perform electrochemical polishing at temperatures in the range of -30°C (Dumitraschkewitz et al., 2019; Lefebvre et al., 2002). Ongoing worldwide efforts aimed at pushing the development of "cryo-APT" for specimen preparation and specimen transfer (Perea et al., 2017; McCarroll et al., 2020; Stephenson et al., 2018) might help taking a step forward in the feasibility of APT analysis of layers of liquids or liquids embedded into a hard structure – see for instance the recent report of APT analysis of an hydrated glass by (Schreiber et al., 2018) – and the associated solid-liquid interface. Efforts involving graphene encapsulation akin to a liquid cell for TEM have been reported by the group at Monash–Deakin (Qiu, Garg, et al., 2020; Qiu, Zheng, et al., 2020). The liquid is sandwiched between a single graphene sheet and a sharpened metallic needle. The liquid volume was very small and the freezing was uncontrolled, leaving many questions unanswered as to whether the distribution of elements within the liquid was affected by the freezing process.

To enable precise analyses on larger volumes, which are more relevant to nanoparticles or biological systems, advanced strategies for specimen preparation and systematic studies are necessary to assess the performance limits of cryo-APT. The development of these methods could, in part, borrow from the biological sciences – for instance cryo-scanning electron microscopy

(SEM) and cryo-FIB are more common in biology, but are now emerging in APT (Lilensten & Gault, 2020; Rivas et al., 2020; Chang et al., 2019; Schreiber et al., 2018). Such protocols could unlock the potential for APT of carbon-based materials, and even go beyond the studies scattered across the literature (Panitz, 1982; Narayan et al., 2012; Prosa et al., 2010; Perea et al., 2016; Rusitzka et al., 2018; Gault et al., 2009; Nickerson et al., 2015; Eder et al., 2017). In addition, they could offer an opportunity to study active materials, e.g. catalysts, in their native or *in operando* environment. Typically, conventional FIB-based preparation methods remove samples and embed in them in a foreign medium, which may alter or damage them (Felfer et al., 2015; El-Zoka et al., 2017; Kim et al., 2018, 2019). Near-atomic scale analysis of 'bulk' ice may also lead to insights into the behaviour of solutes in solution for instance, including distribution of impurities or segregation in natural ice.

Here, we discuss the details of two recently developed specimen preparation strategies that have enabled the APT analysis of frozen water, frozen water-based solutions, and embedded nanostructures. The first approach utilizes liquid-nitrogen (LN2) plunge-freezing a flat nanoporous gold (NPG) film previously immersed in water.. The ice film is subsequently turned into a needle using a moat-type approach (Halpin et al., 2019). The second approach utilizes a commercial plunge-freezing device leading to the likely formation of amorphous ice from a droplet deposited with a micro-pipette on top of a wire-shaped blank. The droplet is then reduced down into a sharp needle by annular milling. Both protocols, i.e. on the flat substrate or the wire, then involve cryo-vacuum transfer into an atom probe for analysis. APT data from both approaches are shown and briefly compared. The challenges inherent to these two approaches are also discussed, as well as some of the key problems that will need a community effort to be tackled, to fully enable routine APT analysis of frozen liquids.

# 2 Porous-Flat-substrate-based preparation

## 2.1 Overview

This first approach makes use of the infrastructure of the Laplace project detailed by Stephenson et al. (Stephenson et al., 2018). As summarised in Figure 1, the complete workflow from the freezing to the APT analysis without breaking ultra-high vacuum and cryogenic temperature during transferring involves four main steps, all described in the sections below.

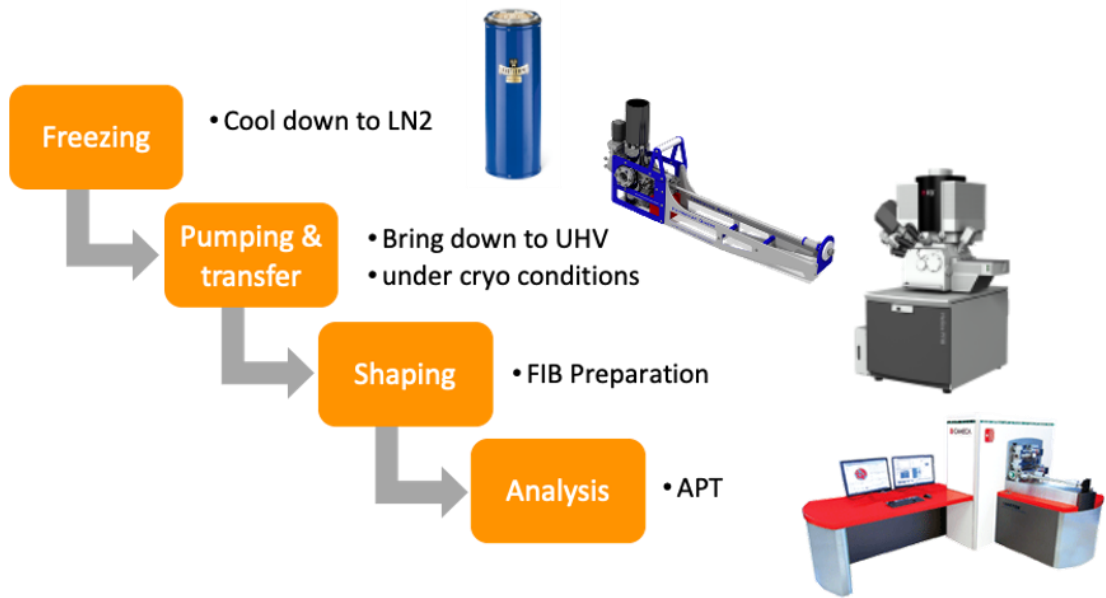

**Figure 1: workflow adopted for the analysis of frozen liquids on flat substrates.**

## 2.2 Formation of the nanoporous gold

Nanoporous gold (NPG) is formed by the dealloying of Ag in acidic, oxidising conditions from a solid solution of AgAu (Erlebacher et al., 2012; Newman, 2010). The selective dissolution of Ag occurs concurrently with the surface diffusion of Au, leading to the formation of a three-dimensional, open-pore, bicontinuous structure. Pore/ligament sizes could be as fine as 3-20 nm (El-Zoka, Howe, et al., 2018; El-Zoka, Langelier, et al., 2018). NPG exhibits a high surface-area-to-volume ratio and as such has found potential applications in electrochemical sensing and actuation (Xue et al., 2014), as well as in catalysis (Zugic et al., 2017). Noting the considerable hydrophilicity of NPG compared to flat polycrystalline Au (Raspal et al., 2012), and that solutions can penetrate well into the thickness of NPG (El-Zoka et al., 2017) we sought to take advantage of NPG as a substrate for APT analysis of frozen water and salt-water solution. A $Ag_{77}Au_{23}$ foil with a nominal area of ~1 $cm^2$ and a thickness of 150 µm was first mechanically polished, and then annealed for 1h at 900°C in an Ar atmosphere. The foil was immersed for 5 mins in a solution of 65 % nitric acid. The foil was then transferred into heavy water, i.e. $D_2O$ (Sigma-Aldrich, Germany, 99.9 at. % D) in order to stop the dealloying. The foil was then mounted on a commercial clip holder, typically used to hold Si-microtip coupons in the Cameca atom probes. Once mounted, the sample is left immersed in $D_2O$ overnight to avoid drying.

## 2.3 *Manual plunge-freezing and specimen transfer*

The immersed clip-holder was first removed from the water. A paper-fibre cleaning wipe (Kimwipe) was then used to gently blot the surface, leaving behind a thin water film on the surface, thereby avoiding excessively large volumes of ice forming on the foil's surface. The freezing of the water film was achieved by manually plunge freezing the water-bearing NPG foil on the clip-holder into dry liquid $N_2$ for approx. 5 mins, and under a constant flow of dry liquid $N_2$. To avoid frosting, the entire process took place inside a glovebox with the oxygen level and the dew point maintained below 1 ppm and below -99 °C, respectively. The sample holder was then mounted directly into the ultra-high vacuum cryogenic transfer (UHVCT) unit mounted onto the glovebox. The UHVCT is connected through a fast pump down docking station, partly built from commercial components from Ferrovac GmbH, equipped with a primary and turbo-molecular pumping system, as well as a cryo-pump, which can achieve UHV conditions within less than 5 mins. The UHVCT was pre-cooled to LN2 temperature. The infrastructure is referred to as the Laplace Project and is detailed in ref.(Stephenson et al., 2018).

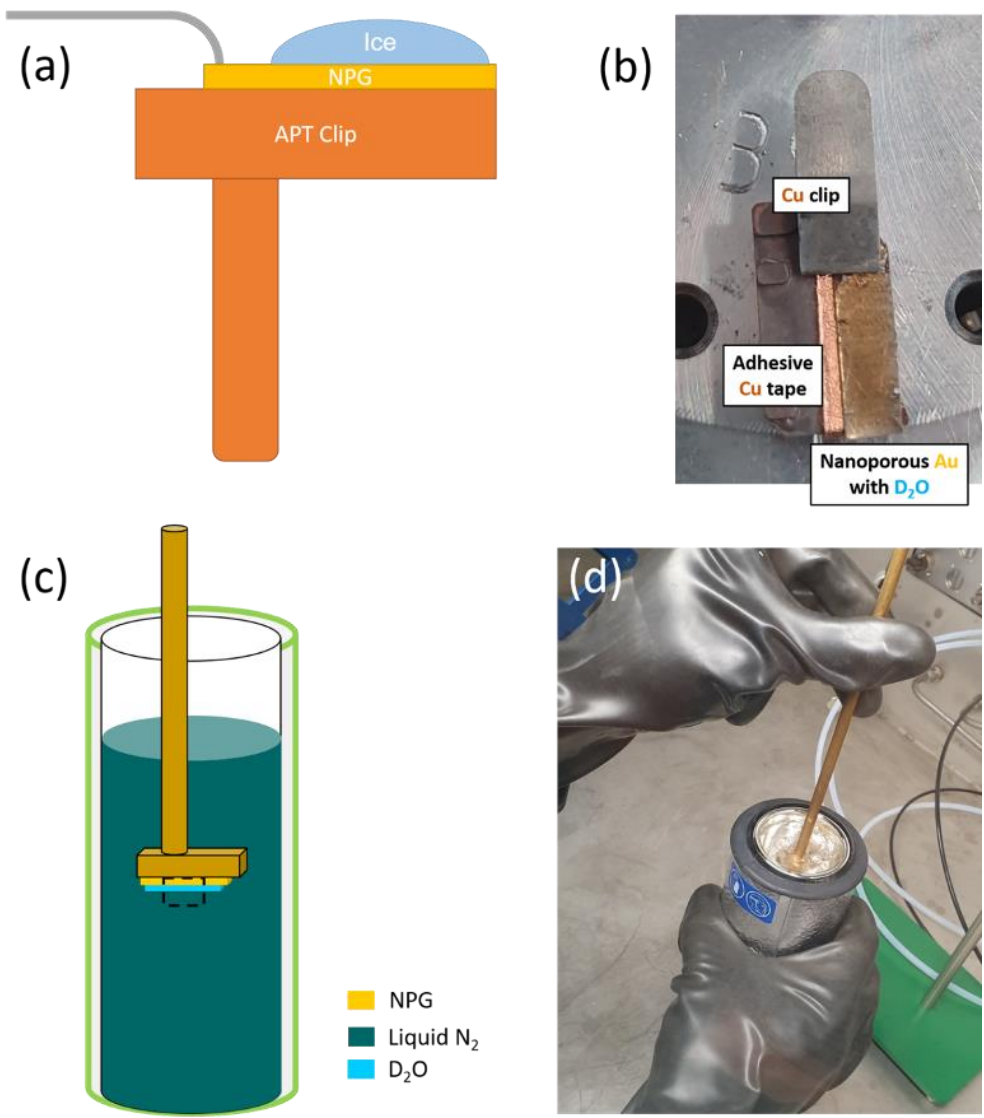

**Figure 2: (a) schematic view of the setup for the preparation on nanoporous gold; (b) photo of the clip-holder and the water-bearing NPG foil; (c-d) schematic and photo of the plunge-freezing setup.**

## 2.4 Cryo-PFIB

The UHVCT is docked onto a dual-beam scanning electron microscope/focused ion beam (SEM/FIB) FEI Helios PFIB (Xe plasma FIB). A custom-designed stage is cooled by a set of copper bands connected to a cold finger fitted onto a dewar filled with LN2, and isolated from the microscope's body by spacers made of polyether ether ketone (PEEK). The stage was designed to accommodate a commercial Cameca APT sample-holder, known and referred to as a puck. Following transfer, we imaged the ice layer by SEM and performed cross-section imaging to visualise the details of the ice layer.

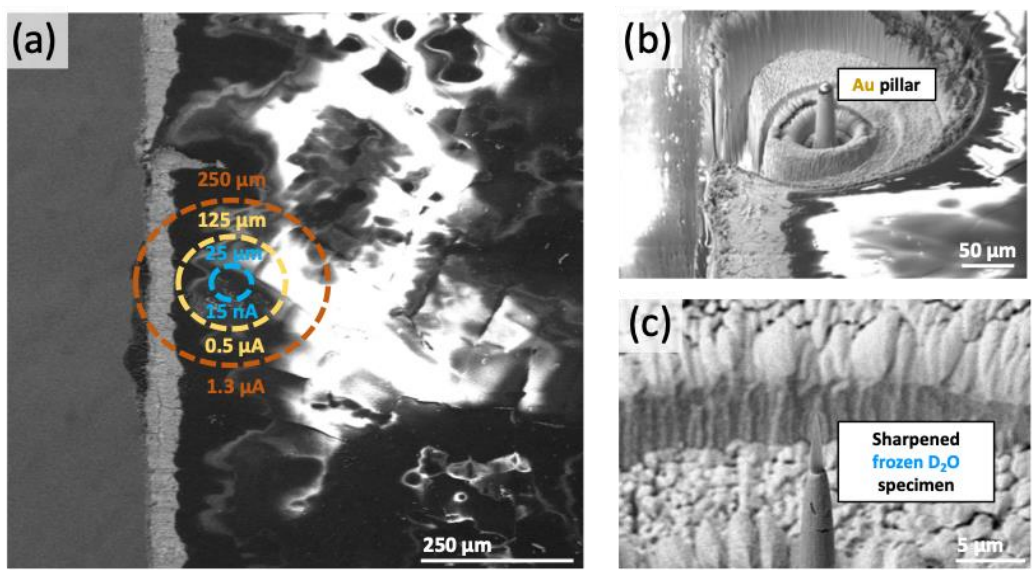

**Figure 3: (a) SEM image of the edge of the water droplet on the NPG and indicative position of the specimen with the outer diameter of the annular pattern and current used to mill the pillar. (b) Au pillar with the ice-layer. (c) sharpened APT specimen of ice on the NPG.**

To obtain specimen suitable for APT analysis, we adapted the protocol of Halpin et al. (Halpin et al., 2019) to prepare a pillar that is accessible to the laser illumination within the atom probe itself. The pillar is shaped into a needle by a series of annular patterns. The Xe ion currents were progressively adjusted from 1.3 µA down to 0.5 µA, and finally 15 nA. The acceleration voltage was maintained at 30kV to form the central pillar on which the final specimen was later prepared. These steps are detailed in Figure 3(a), and an image of the final pillar is shown in Figure 3 (b). No obvious voids or gaps between the ice and the NPG could be seen following FIB-milling. The final sharpening of the specimen was achieved by reducing both the inner and outer diameters of the annular milling pattern and reducing the current progressively from 4 and then to 1 nA. We ensured that the length of the ice layer inside the specimen was consistently below 5 microns. The finalised specimen is shown in Figure 3 (c).

Following shaping of the specimen's, the puck is transferred back into the LN2-cooled UHVCT via a UHV side chamber kept at approx. -160 $^{o}$C. The transfer time is below 15 sec. We detached the UHVCT from the PFIB and mounted it for direct transfer into the buffer chamber of a Cameca LEAP 5000XS, onto a pre-cooled so-called piggyback puck placed into a slot on the carousel insulated from the rest of the microscope by PEEK. The puck is hence passively cooled until transferred into the atom probe analysis chamber.

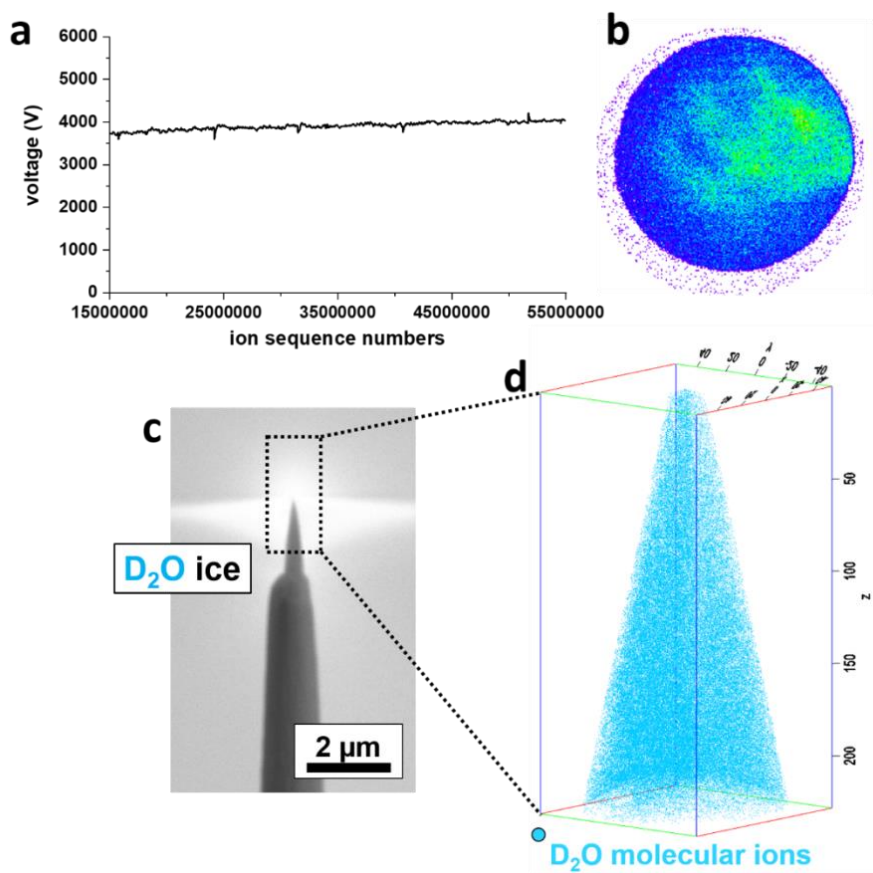

**Figure 4: a-b voltage curve and detector hit map obtained from an ice specimen shown in c. d tomographic reconstruction obtained from this dataset.**

## 2.5 Atom probe tomography

The APT analyses were performed using a Cameca local electrode atom probe (LEAP) 5000 XS (Cameca Instruments, USA). The data was acquired in laser-pulsing mode with a pulse energy in the range of 20–100 pJ at a base temperature of 70K. The Integrated Visualization and Analysis Software (IVAS) 3.8.4 was used for data reconstruction and analysis. Custom MATLAB routines were used for calculating and displaying correlation histograms. Figure 4 is a summary of the main results. The voltage vs. number of acquired ions in Figure 4a, which shows a rather smooth evolution of the voltage, indicative of no major bursts of detection or specimen fracture. Figure 4b is the corresponding detector impact histogram, with some regions of high impact density (in green) and low impact density (in blue). Figure 4c–d respectively show a scanning electron micrograph of the finalised specimen and the corresponding reconstruction in which all D2O molecular ions are displayed as individual blue dots. The default reconstruction parameters were used and the atomic volume of oxygen was used as a basis for the depth increment calculation (Gault et al., 2011).

# 3 Preparation on wires

## 3.1 Overview

The wire approach presented in this section borrows even more from the workflows for cryogenic imaging of biological samples. A suite of commercial equipment from Leica was used: a vacuum-cryo-manipulation system (EM VCM), a high vacuum-coater (EM ACE600) and a cryo-vacuum shuttle (EM VCT500). The used FIB Instrument (Scios FEI) was equipped with a dedicated cryo-stage to keep samples at a minimum temperature of 130 K. The stage is connected by copper stripes to a LN2-Dewar mounted at a custom-made side flange, to keep the whole cooling process vibration free. The stage was further modified to accept a standard APT sample holder. The FIB is equipped with an EasyLift$^{TM}$ micromanipulator, which is limiting the preparation possibilities at cryogenic temperatures, since it is operating at room temperature. Therefore, standard lift-out procedure was not readily possible, conversely to the protocol proposed by Schreiber et al for instance (Schreiber et al., 2018). Figure 5 summarises the complete workflow from the freezing to the APT analysis. The main difference in this workflow compared to the previous one, is the use of pre-sharpened wire blanks with a flat, rough surface.

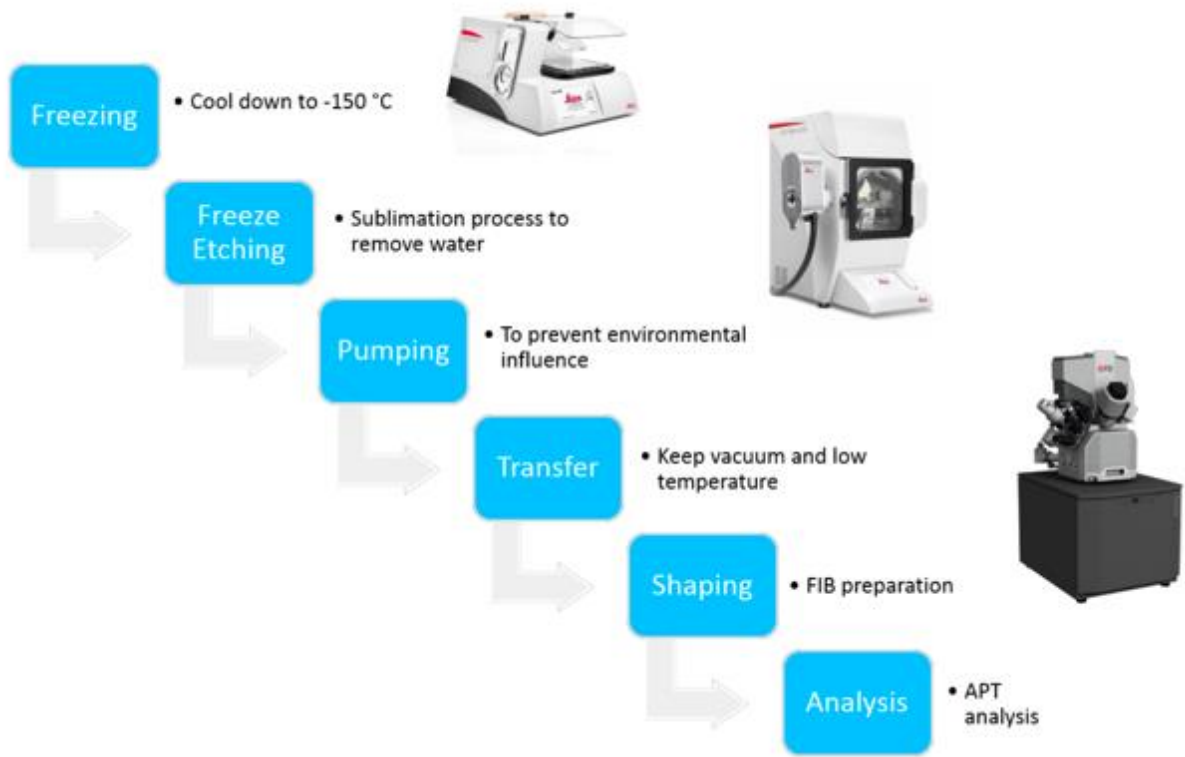

**Figure 5: workflow adopted for the analysis of frozen liquids on wire blanks.**

*3.2 Dipping technique for viscous liquids on pre-shaped wires*

Tungsten (W) was selected as a substrate material, owing, in part, to its wide availability in most laboratories as it is a well-established test material for APT. Following the common specimen preparation technique for sputter deposited layers, where pre-shaped tungsten tips are used as template, we attempted making needle-shaped W specimen suitable for APT analysis. The needles were prepared by dipping a pure W-wire into a 2 M NaOH solution, while applying an AC voltage of 3–7 V, which lead to the formation of a neck in the middle of the wire. The final shape was achieved by switching to a pulsed voltage of 2–3 V up to the point where the wire breaks into two parts. Needles with a typical end radius below 100 nm, were subsequently dipped in honey and imaged in the SEM upon freezing to -150°C. The fluid, despite its high viscosity, did not stick to the needle's end but migrated along the shank. This might be attributed to the low radius of curvature of the tip apex, not compatible with the surface tension of the liquid and the low contact angle, as displayed in Figure 6.

In a next step the apex of the tip was removed by subsequent FIB cutting. As result a tungsten post with a diameter of approx. 2 µm was fabricated, turning it into a flat top blank. Yet the result was the same and the drop of fluid did not adhere on top of the post but further down the shank, making this approach unfit for making APT specimens from frozen fluids.

Our goal was to produce droplets as small as possible, since any increase in diameter of the droplet would result in a dramatic increase in later FIB annular milling processes to produce a suitable tip. Consequently, we increased the post diameter by using commercially available tungsten wires (20 µm to 100 µm). First positive results of droplets sticking to the flat surface of the post were achieved using 75 µm thick wires. Needless to mention, that the production of flat post with 75 µm diameter using the FIB was a time-consuming procedure and not applicable for fast sample throughput. To get access to a fast-available substrate we decided to omit any pre preparation using the FIB instrument.

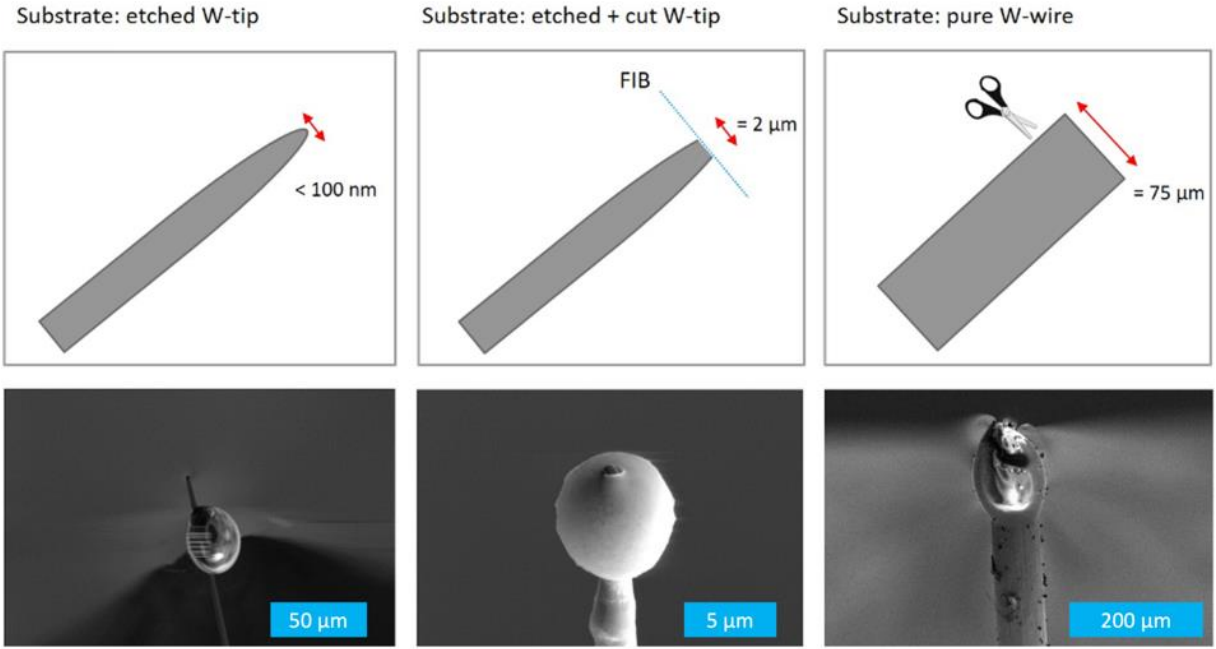

**Figure 6: Representative dipping results for W wires with different sample geometries in honey.**

A tungsten wire, 75 µm in diameter, was first cooled to liquid nitrogen temperature (-196 °C). At such low temperature, for tungsten in a body-centered cubic crystal structure, brittle fracture with small plastic deformation and neck formation can be expected. The wire was fractured by applying a tensile force with two pincers which were also cooled down before. The fractured surface is, as illustrated in Figure 7 (a-b), about 50 µm in diameter due to some necking, very rough, with cracks and crevices oriented along the wire's main axis (Figure 7 (b)), and hence offering a much wider surface area and enhancing the likelihood that enough fluid would remain on the blank's top surface. After creation of the rough surface, the respective tungsten post is dipped into the high viscosity liquid. The post with the formed droplet on its top is plunged frozen in a LN2 bath. Adjacent transfer of the cooled sample into the Leica VCT 500 Shuttle and suitable pumping sequences allow to follow the further preparation route.

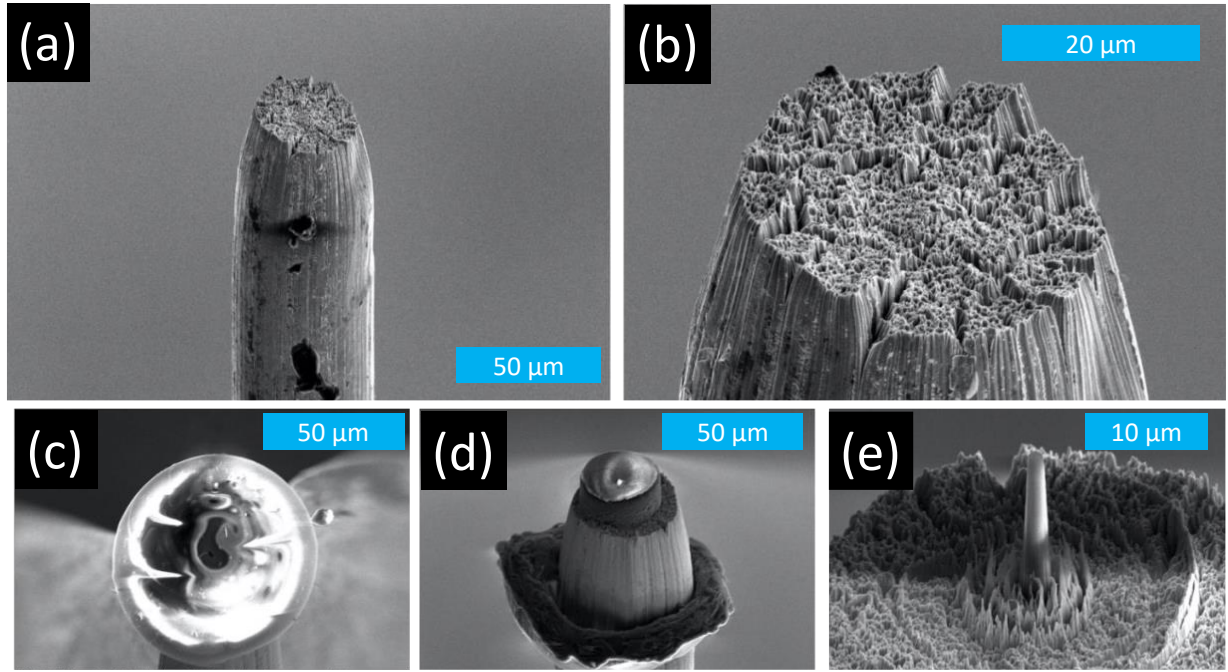

**Figure 7:** scanning electron micrographs (a) of the fractured wired, and (b) close-up on the fractured surface itself; (c) honey drop located at the blank's tip, and (d) the metal-honey interface after the initial annular milling, and (d) pillar ready for final shaping into an APT specimen.

### 3.3 Freezing of water droplets

As shown in the previous subsection, the dipping technique using flat tungsten posts works well for high viscosity fluids such as honey. Using this method, it was possible to freeze droplets of reasonable size on top of the post. But for liquids with low viscosity (water, heavy water and NaCl aqueous solutions), all attempts failed to create a similar droplet on the surface. Only a very thin layer of a few tens of nanometer was detectable, stemming most likely from moisture condensed on the cold wire during transfer.

So, for low viscosity liquids, the protocol changed in that way, that the post is first mounted on a sample holder and placed into a LN2 bath (Leica EM VCM). The nitrogen atmosphere above the liquid protects the wire against condensation of moisture from the surrounding atmosphere. A water droplet is dipped with a micropipette onto the precooled tungsten post, which is stored in the LN2 bath (Figure 8). The droplet forming at the nozzle of the micropipette is dipped onto the pre-cooled tungsten post in a free fall mode, so that it does not touch the metal surface before it constricted. Falling below the critical distance can cause the liquid to stick to shaft of the post. If possible, only one dipping process should be done, because each additional dipping creates an interface between the two droplets, which is a typical point at which the upper material breaks off during the milling process.

We used pure water, deionized and filtered through a commercial Millipore Milli-Q system (purity characterized in terms of resistivity $\rho > 18$ M$\Omega \cdot$ cm). The high purity should prevent additional peaks from appearing in the mass spectrum. The frozen droplet is 100-200 µm long and 150-250 µm in diameter, and can subsequently be turned into a needle-shaped specimen suitable for APT. This approach is summarized in Figure 8.

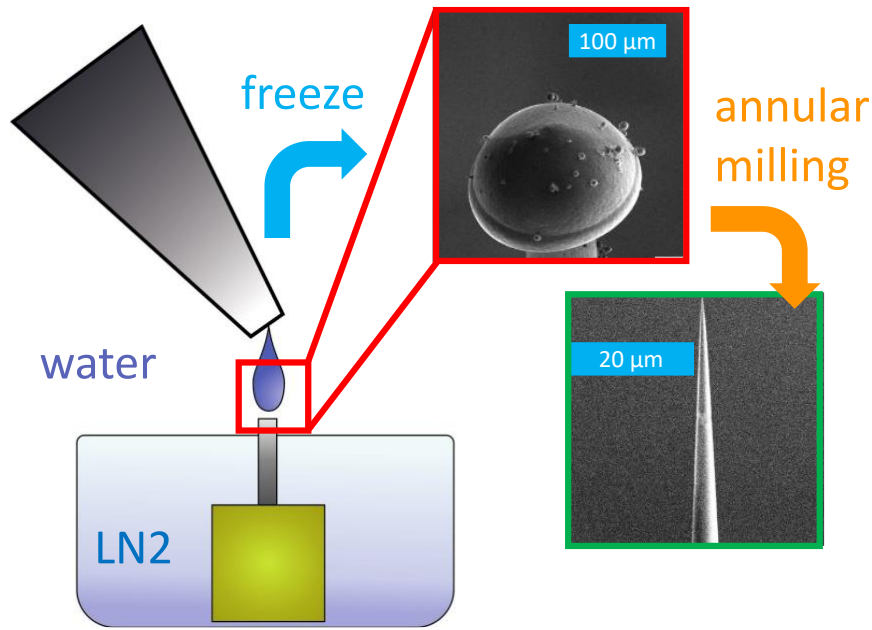

**Figure 8: summary of the approach for making frozen droplets and shaping them as needles for APT.**

Hereupon, the sample holder is transferred into the cooled body of the modified transfer shuttle VCT500 from Leica (T= -164 °C), which is initially pumped by an oil free scroll pump to a pressure of $6 \cdot 10^{-1}$ mbar. To allow the transfer into the FIB the vacuum has to be improved by a further intermediate step. The shuttle is attached to the high vacuum-coater (Leica EM ACE600 Fig. 4)), in which a freeze etching (FE) process can be carried out to remove ice crystals which are formed by contact with air. By heating up the sample very precisely to a temperature of -90 °C and a pressure of $9 \cdot 10^{-7}$ mbar for 30 minutes, a sublimation process from solid ice to vaporous ice occurs, which allows the controlled removal of condensed ice from the environment. On the other hand, the Leica EM ACE600 high vacuum coater is necessary to improve the vacuum conditions inside the VCT500 shuttle into a range of $10^{-4}$ mbar, which is necessary precondition for the transfer into the FIB. If the pressure of the shuttle is not sufficient the vacuum safety measures of the instrument will be engaged.

*3.4 FIB milling*

Following freeze etching, the VCT500 is used to transfer the frozen sample into a FEI Scios SEM/FIB via a dedicated load lock. The SEM/FIB is equipped with a custom-designed cryo-stage. The stage is cooled by copper bands connected to a Dewar willed with $LN_2$, allowing to reach a temperature down to approx. -150 °C. A cryo-shield is also connected to the Dewar, to act as a cold trap and help avoid re-deposition on the sample during the preparation of the needle-shaped specimen. SEM imaging was typically performed with using low acceleration voltage, i.e. 5 kV, and beam current 25 pA and only by taking snapshots of high scanning speed to prevent melting of the sample by electron bombardment. Live imaging of the sample or intense focusing results in fast melting of the adhered droplet. Following previously reported preparation protocols, the sample is tilted to 52° for annular milling (Prosa & Larson, 2017).

An annular pattern was used, which mills from the outer to inner part. In a first step, all protruding parts of the droplet (Ø 150-250 µm) are cut away at 30kV acceleration voltage and a very high ion beam current, approx. 50 nA, until the edges of the 50 µm diameter tungsten substrate are visible. The ion beam current is then progressively reduced to 0.1 pA along with the inner diameter of the annular milling pattern, down to 600 nm. The precise current and dimensions of the annular pattern are reported in

. SEM imaging is performed intermittently to monitor the progress of the specimen's shaping, as showcase in Figure 9. The finished specimen, with an end radius in the range of below 100nm is finally transferred into the APT via the VCT500. The preparation time always depends on the material and of course on the size of the drop and the thickness of the material on the post, but it can be said that the average preparation time is about 3–5 hours per specimen. The reproducibility of the approach is highlighted by the six different specimens shown in Figure 10a–f.

**Table 1: currents and size of the annular pattern used through the FIB-preparation process.**

| Current | Outer Diameter | Inner Diameter | Z- Direction | Current | Outer Diameter | Inner Diameter | Z-Direction |
|---|---|---|---|---|---|---|---|
| **50 nA** | 230-90 µm | 180-40 µm | 50 µm | **1 nA** | 30 µm | 4 µm | 5 µm |
| **30 nA** | 50 µm | 30 µm | 50 µm | **0.3 nA** | 20 µm | 2 µm | 5 µm |
| **15 nA** | 50 µm | 20 µm | 10 µm | **100 pA** | 15 µm | 1 µm | 1 µm |
| **7 nA** | 50 µm | 10 µm | 10 µm | **49 pA** | 10 µm | 600 nm | 1 µm |
| **5 nA** | 30 µm | 8 µm | 10 µm | **10 pA** | 10 µm | 300 nm | 1 µm |
| **3 nA** | 30 µm | 6 µm | 5 µm | **10 pA** | 10 µm | 150 nm | 1 µm |

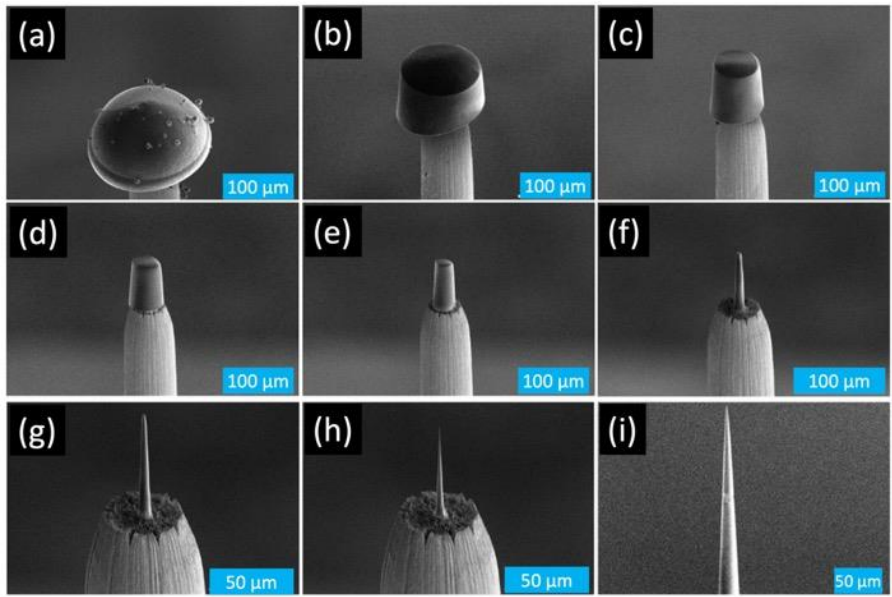

**Figure 9: a–i Scanning electron micrographs showing the various steps of the preparation process, from the large drop to the needle-shaped specimen suitable for APT.**

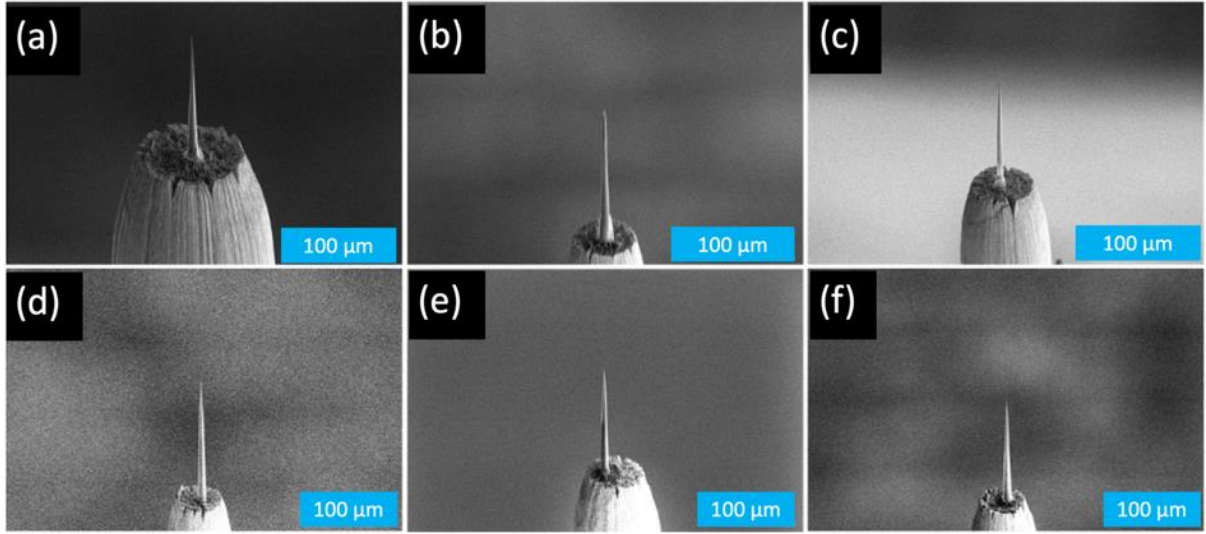

**Figure 10: a-f six different needles presented in order to show the reproducibility of the length and the shape of the specimen.**

### 3.5 Atom probe analysis

APT analysis was performed on a custom-made atom probe (Schlesiger et al., 2010), equipped with a ClarkMXR laser system with a fundamental wavelength of 1030nm – similar to the instrument commercialized by Inspico. By usage of the second and third harmonics the laser wave length can be varied between 515nm and 355 nm. The system has been modified to accommodate the VCT500 and to enable a fast exchange of the cryogenic samples. The data from the ice specimens was acquired in laser-pulsing mode, with 250 fs laser pulses at a wavelength of 355 nm, focused to a spot size of approx. 50 µm and a repetition rate of 100kHz. The detector comprises microchannel plates with an open area of 50% and a delay-line detector with a diameter of 120mm. The sample for APT analysis was cooled down to 63K. Volume reconstruction was performed following the original point projection method by Bas et al. (Bas et al., 1995) using the specimen radius derived from the method proposed by (Jeske & Schmitz, 2001). The obtained datasets were analyzed using Inspico's Scito software package. SEM pictures were used to determine the initial specimen radius, but high-resolution imaging could not be performed due to possible melting of the sample. Therefore, is the respective value afflicted with a rather large error.

# 4 Discussion

## 4.1 Specimen preparation strategy

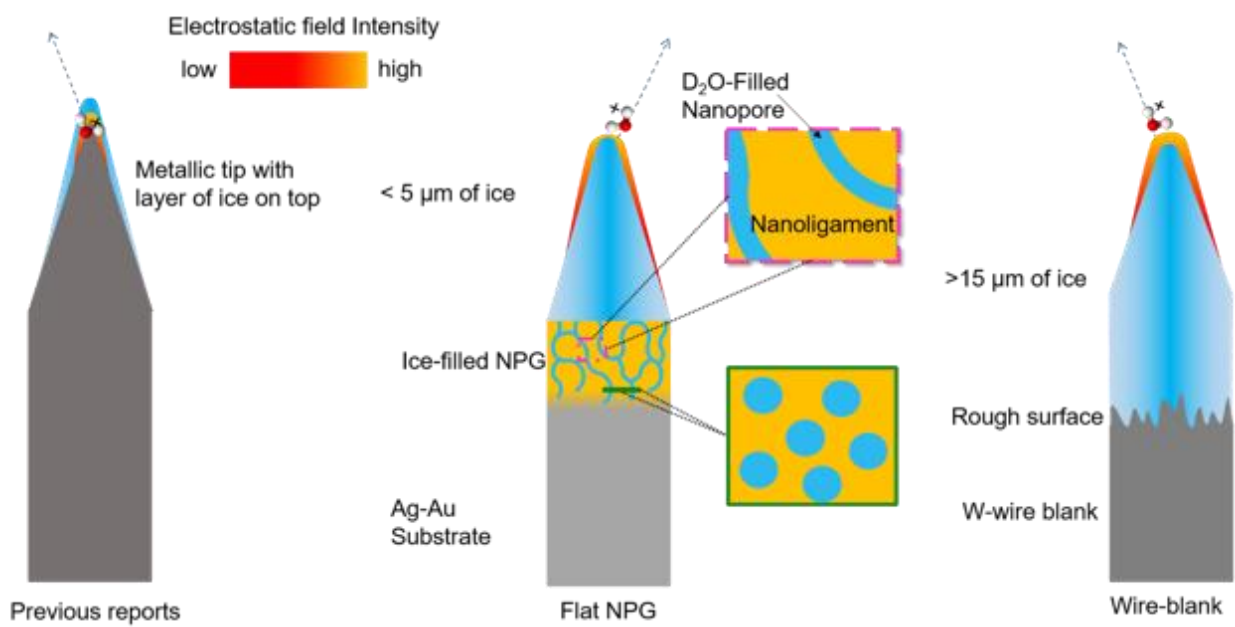

**Figure 11: comparison of the two specimen preparation strategies in comparison to previous reports.**

Both approaches have yielded results, and will, in principle, keep doing so. As preliminary efforts, these protocols will surely be refined over time. Here the field evaporation takes place through a thick layer of ice formed on a flat substrate, which contrasts with previous reports: where a thin layer of ice was formed on a sharpened specimen (Stuve, 2012; Tsong, 1985; Pinkerton et al., 1999), as summarised in Figure 11.

We have attempted a range of other possibilities before selecting NPG as a substrate, including a silicon wafer and a commercial microtip coupon, but the hydrophobicity of the substrate made the preparation of suitable specimens almost impossible. The attempts with the thin wire blank discussed above show that the task to find a suitable substrate is arduous. The hydrophilicity and low contact angle of the water droplet on the NPG facilitated the preparation of the specimens by the moat-type approach. The continuous network of pores provided strength to the ice-substrate interface, allowing for analysis of the interface. This is, in principle, also the case for the rough fractured surface of the blank itself, yet no results were obtained since the ice layer in this case was above 15 µm. The thickness of the ice layer in this latter case could make it easier to embed larger structures – an E.coli bacterium is approx. 1 µm in diameter, a typical red blood cell is 2–5 µm, and a human cell 10–20 µm. In addition, the evaporation field of water appears to be rather low, and the contrast with W might prevent establishing a specimen-shape suitable for the field evaporation of the W substrate itself.

The flat substrate approach could in principle allow for depositing solutions containing a range of different solutes or objects. A key limitation of this approach however is the cooling process. In order to obtain a fully amorphous layer of ice, it is necessary to reach a cooling rate of $10^6$ K/s or higher at ambient pressure (Moore & Molinero, 2011), to reach the glass transition temperature (136 K). Estimations of the cooling rate with the current setup indicate that it is 2 orders of magnitude too slow at least. The large volume of water, and of the bulky metallic substrate underneath, likely makes this harder to control, and maybe impossible to reach the cooling rates necessary for vitrification.

The blank approach presented herein should facilitate the vitrification, as it makes use of commercial devices designed for the cryo-preparation of biological samples, adapted in this case for APT. The crystalline or amorphous nature of the ice analysed here was not checked. It is likely that this can only be confirmed by TEM. An APT specimen made on a wire blank is a geometry perfectly suitable for TEM imaging, such an additional step would make these cryo-protocols even more complex, and could very well lower yield of successful experiments. In addition, there is a possibility to perform either electron backscattered diffraction (EBSD) on the ice layer (PRIOR et al., 2015) or transmission Kikuchi diffraction to verify crystallinity of samples. The former may be easier to apply with the flat substrate method, provided that the appropriate tilt is achievable using the cryo-stage in the SEM. While the wire blank approach is more versatile, it may not easily allow for site specific preparation, whereas the flat substrate should allow for making the moat, pillar and final needle-shaped specimen from various regions of the substrate, given that the area of choice is close to the substrate's edge.

There are many options that could be explored in the future – changing the pore size, the nature of the substrate etc. and even for crystalline ice, there lies ahead many opportunities for key studies of solute segregation and redistribution during solidification. Using a water-based solution as a carry medium for embedding nanostructures is also now possible, and crystalline ice may help locate these nanoparticles close to the substrate's surface, making their preparation somehow simpler. The use of the flat substrate may also facilitate the performance of cryo-liftout, to select specific regions in large samples to analyse (PARMENTER & NIZAMUDEEN, 2020).

## 4.2 Data comparison and similarities

The two approaches enable the analysis of layers of ice of ~ 2–80 µm in thickness. Atom probe analysis from insulating materials has now become almost routine, and it was proposed that the electrostatic field and photon absorption by the specimen are enabled by the field-induced bending of the electronic bands combined with surface defect states (Silaeva et al., 2014; Kelly et al., 2014). Whether similar processes take place in ice cannot be concluded at this stage, since the electric conduction in ice is not expected to involve electrons but protons. Nevertheless, APT data was acquired from micron-thick layers of frozen water and solutions.

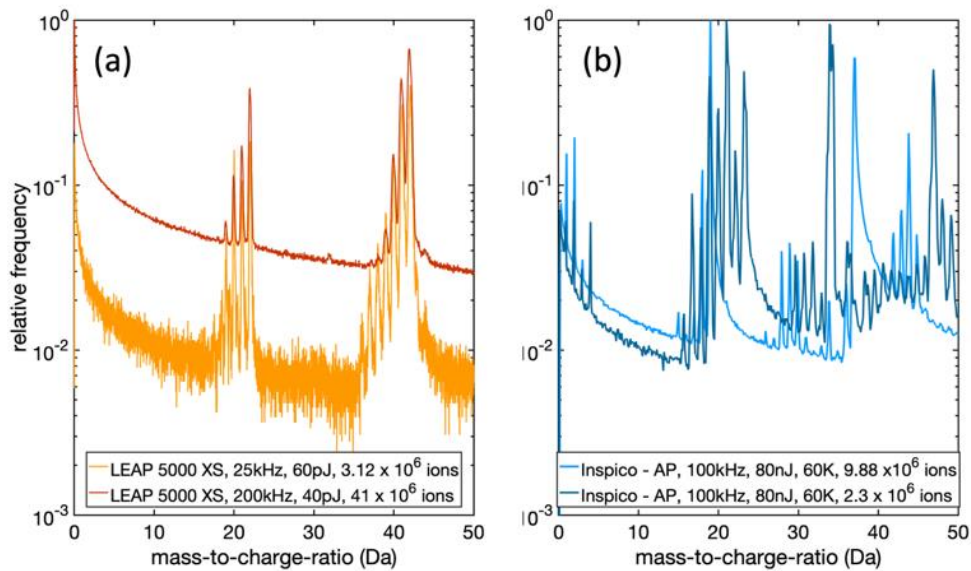

**Figure 12: mass spectra for the analysis of D$_2$O ice obtained on the flat substrate (orange, data acquired on a Cameca LEAP) and both H$_2$O and D2O ice obtained on a wire blank (blue, data acquired on an Inspico-AP).**

Examples of mass spectra are shown in **Figure 12** for two datasets obtained from the flat substrate on the LEAP 5000XS for two laser pulse energies and repetition rates in orange-red (**Figure 12**a), and in blue is an example obtained on a wire blank on the Inspico atom probe (**Figure 12**b). The mass spectra were normalised with respect to their relative frequency value. In both cases, a series of peaks correspond to various molecular ions containing a number of O atoms with a combination of D and H atoms. All molecular ions appear to be protonated, and mostly bearing a 1+ charge. This strongly supports the idea that protons are the only relevant charge carrier. The formation of metastable molecular ions from a combination of water molecules was discussed from a theoretical perspective by Karahka & Kreuzer (Karahka & Kreuzer, 2011) and is consistent with previous experimental reports (Tsong, 1985). For data from the LEAP, **Figure 12**a, the drop in the level of background from 60pJ@25kHz and 40pJ@200kHz can be ascribed primarily to lower electrostatic field leading to less evaporation at the DC field in between the thermal pulses, as discussed in detail in ref. (El-Zoka et al., 2020).

The data obtained for and $H_2O$ and $D_2O$ under similar conditions and on the same instrument, **Figure 12**b, highlight the complexity in the formation of the molecular pattern associated to the dissociation process, as studied in part by Schwarz et al.(Schwarz et al., 2020). The increased atomic weight for deuterium-containing molecular ions will make them travel more slowly in the early stage of their flight, where post-ionisation (Kingham, 1982) and possible field-induced dissociations (David Zanuttini et al., 2017) are more likely.

Interesting points arise from the comparison of the data from the two instruments and specimen preparation strategies. For now, it is simply not possible to completely disentangle the origins of these differences. The levels of background are comparable for the light orange and light blue mass spectra. The relative amplitude of the smaller clusters in the case of the lower laser pulse energy is coherent with the increase in the electrostatic field necessary to maintain the detection rate at 5 ions per 1000 pulses on average. However, the sets of peaks for lighter molecular ions appear to be more abundant in the data from the Inspico instrument, **Figure 12**b. A comparison between the patterns formed for $D_2O$ on the two instruments, **Figure 12** a and b, highlights that the data acquired from the flat substrate on the LEAP is at a lower overall field compared to the wire-blank in the Inspico instrument, as no atomic H or D arising from dissociations are observed in the orange mass spectra. This is also consistent with observations on many other elements with a tendency to form molecular ions during APT analysis (Müller, Saxey, et al., 2011; Mancini et al., 2014; Santhanagopalan et al., 2015).

In addition, the set of peaks around 44Da appear in **Figure 12**b and were attributed to (HOOOH)(H$_3$O)$_2^{2+}$, i.e. a doubly charged ion (Schwarz et al., 2020), using simulations based on density functional theory. There is also a possibility that it could be Ga(H3O)$^{2+}$, as Ga was used for the specimen preparation. Either way, the appearance of doubly-charged ions suggests higher electrostatic field conditions. The laser fluence in the case of the data from the flat substrate analysed in the LEAP 5000XS in the range of 5–8 J/m$^2$ while it is approx. 45–50 J/m$^2$ for the data acquired on the wire blank on the Inspico instrument, and there are three orders of magnitude difference in the peak intensity. These illumination conditions should in principle lead to the data acquired on the Inspico instrument to be under much lower electrostatic field conditions, which is opposite to what is actually observed. Finally, the shape of the mass peaks from the wire blank also suggests that there are significant heat tails, whereas the data from the flat substrate lead to peaks that are almost symmetrical with only little evidence of heat tails. In such a case, the thickness of the ice layer, i.e. 80 µm vs. 5 µm, may be contributing since heat needs to be dissipated through the ice.

The mass spectra in Figure 13 clearly show that the fragmentation pattern, even for supposedly simple molecular liquids like water, are quite complex. Understanding the atomic composition of the peaks is crucial for the reconstruction of the 3D structure, this understanding can be aided by *ab initio* simulations, which determine the stability of possible fragments (Schwarz et al., 2020), model the fragmentation process itself (Nickerson et al., 2015; Gault et al., 2016; Zanuttini et al., 2018; D. Zanuttini et al., 2017) and could also shed light on segregation and other structural changes of frozen water under high electric fields.

## 4.3 Is controlled field evaporation from water achievable?

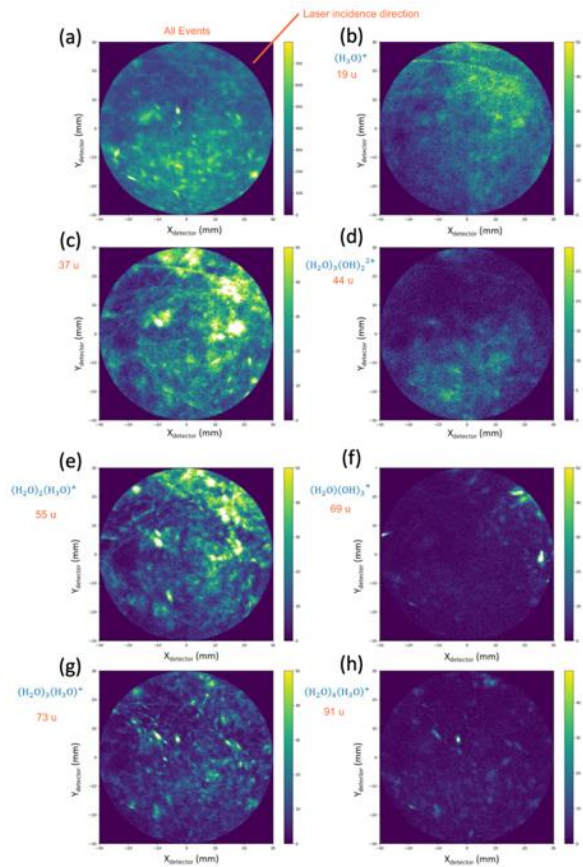

**Figure 13: a-h Detector density maps for the data from the Inspico instrument reported in Figure 10 for the different molecular ions detected (Schwarz et al., 2020).**

Is it possible to reach field evaporation conditions for a frozen liquid that would be comparable to that of solid materials? or are the ions detected from frozen liquids originating from uncontrolled migration/sublimation/ionisation? This question arises as the stability of liquids under UHV conditions is very sensitive to the temperature (Gerstl et al., 2017), and the localized heating associated with laser illumination could lead to localized melting and subsequent sublimation. Answering this question is crucial for any further data reconstruction and processing, since algorithms used for reconstruction typically rely on undisturbed and controlled field evaporation process in order to calculate the third dimension. The uneven distribution of hits on the detector, as seen in Figure 4b, and Figure 13, suggests a diffusion of partially-charged molecules at the specimen surface before field evaporation. These patterns are very similar to the high-temperature field desorption of ice reported in (Stintz & Panitz, 1992).

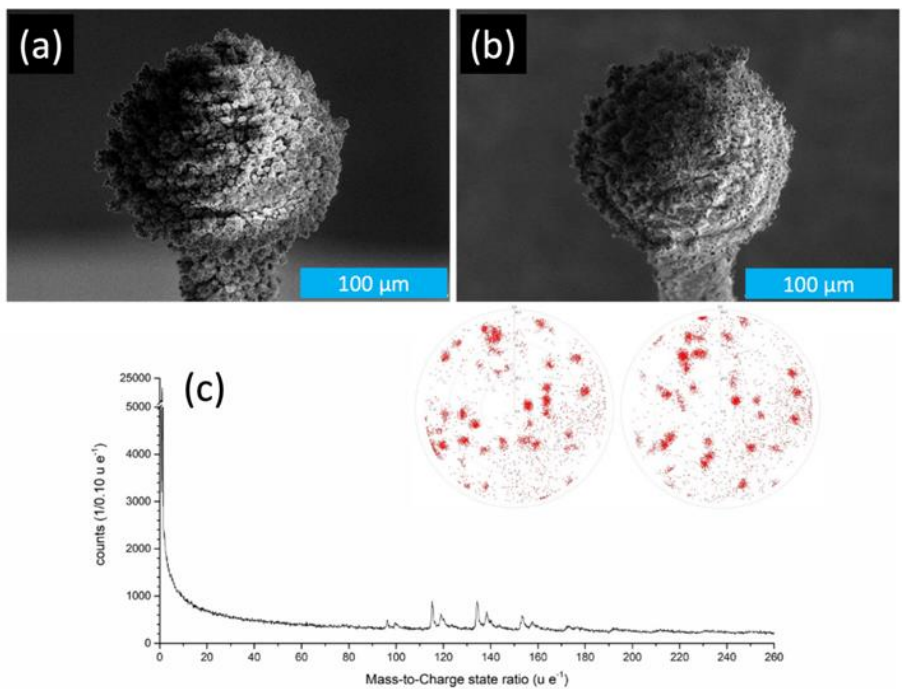

**Figure 14: a) frozen pure mili-Q water droplet without the freeze-etching process before and in b) after the atom probe measurement. c) mass spectrum using high intensity laser illumination and, inset, detector impact maps at two different times during data acquisition.**

To cast light on this problem, we conducted experiments with as-deposited droplets. As shown in Figure 14, the water droplet as well as the tungsten post are decorated with numerous micron-scale ice crystals using moisture condensed during the transfer process. Skipping the sharpening process, the sample was transferred into the atom probe. The laser focus was adjusted to the apex of the droplet with an energy of 15 mW at 100kHz repetition rate on the Inspico instrument. By applying a high voltage of 15 kV, a mostly continuous DC evaporation signal was recorded. From the comparison of Figure 14a, before the measurement, and Figure 14b, following the measurement, it is visible that the most protruding ice crystals at the droplet's surface shrunk significantly, and so did those located on the tungsten wire shank. For very high laser intensities around 40-50 mW, a faint signal above the noise level for large clusters is recognisable, as shown in Figure 14c, but most of the ion count detected is uncorrelated with respect to the laser pulses. In the detector view, small local clusters of evaporation are visible but no homogenous desorption map is obtained. Although the overall radius of curvature of the droplet is in the range of 100 µm, i.e. even at 15kV the intensity of the electrostatic field will be low, the local sharp curvature of the ice crystals will enable continuous field evaporation.

This experiment was repeated with a crystal-free frozen droplet as shown in Figure 15. The surface is much smoother and, as expected, no signal could be detected as reported before using the conventional AP protocol. The laser beam was then moved from the apex of the droplet several microns towards its centre, and with a very high intensity (50 mW at 100 kHz ). As visible in Figure 15b, the surface was modified by the high input of energy, i.e. heat, from the laser. A network of cracks appeared, which, interestingly, coincided with the detection of mass peaks in the mass spectrum that can be attributed to protonated water clusters, Figure 15c. The thermal tails are however very intense, and the collected data shows no homogenous distribution on the detector map, Figure 15c, and only small datasets were collected. However, it shows that the large amount of heat introduced into the frozen water leads to modification of the sample shape and overall structure. Such analysis conditions should hence be avoided.

As a result, to limit DC evaporation and surface modification, we took care to only leave a single frozen water tip on the tungsten post. For the laser energy, the lowest repetition rate and lowest pulse energies for this laser spot size are chosen. The absorption coefficient for ice in the UV and visible spectrum is generally very low, and only a fraction of the net laser intensity is absorbed. Yet the localised heating clearly affects the specimen's shape in a way that has rarely been reported for APT of e.g. metals or ceramics (which is not to say that it does not occur). This striking difference may be intimately coupled to the charge and heat transport mechanisms. Good electrical conductors are good thermal conductors, via electrons, when they exhibit a continuous density of states at the Fermi level, a thermal intra-band carrier creation/annihilation dynamics, and a quasi-ballistic short-range transport. Under these general conditions, "hot carriers" will be generated in areas of high temperature and may thermalize in cooler regions after diffusion. This is certainly the case for metals (Houard et al., 2011), and most likely also applies to semiconductors and insulators in the band-bending scenario, in which the valence band dips below the Fermi level near the surface and becomes partially depleted. In ice, however, charge transport occurs via proton-hopping along existing hydrogen bond networks in a Grotthuss-like mechanism (Park et al., 2014). As these protons neither stem from thermal processes, nor conserve the energy along the hopping path, they cannot drag excess heat with them. The bonding of a proton to H2O is an exothermic reaction, so highly localised heating associated to this process could potentially play on role on the detachment of charged surface species.

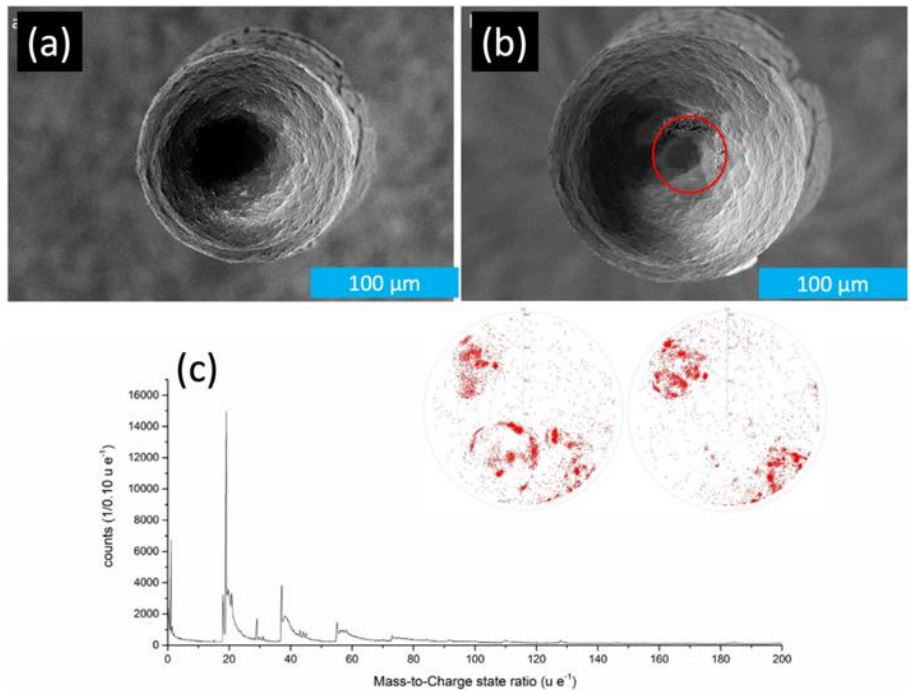

**Figure 15: a-b Water droplet without protruding ice crystals before and after measurement in the atom probe. c mass spectrum using high intensity laser illumination and, inset, detector impact map at two different times during data acquisition. c Corresponding mass spectrum and ion impacts on the detector.**

## *4.4 Open questions and future challenges*

### 4.4.1 Field evaporation and molecular ions

Although a set of preliminary experiments prove the feasibility of analysing ice, there are many open questions since little is known of the field evaporation behaviour of ice, despite the few reports from (Pinkerton et al., 1999; Stuve, 2012) on the very thin ice layers and the thicker layers (Schwarz et al., 2020; El-Zoka et al., 2020). The state-of-the-art is more advanced and efforts on atomistic simulations are ongoing that provide precious insights into the detailed mechanisms of field evaporation in metals (Ashton et al., 2020). Yet, the theory of field evaporation in materials where the conduction is not associated to electrons but to protons (Bilgram & Gränicher, 1974) has rarely been discussed. The formation of water clusters at the surface under the influence of the electric field and the relative stability of various molecular clusters of water was studied experimentally (Stintz & Panitz, 1992; Stuve, 2012) and by *ab initio* methods (Karahka & Kreuzer, 2011). In addition, *ab initio* calculations of the relative stability of molecular ions and their configurations were reported by (Schwarz et al., 2020). A more detailed *ab initio* modelling of the field evaporation process, possibly coupled with proton segregation to the surface, may provide further insight in the cluster formation and its dependence on the applied field and the applied laser power.

An important difference in the experiments reported here, in comparison to previously reported studies is that the highest electrostatic field was at the interface between the sharp metal needle-shaped specimen and the ice, whereas here the highest field is induced by the potential applied at the bottom of a layer of ice that is several microns in thickness. Interestingly, we do not observe significant shift in the mass peak positions that may have indicated a significant voltage drop. Past experiences that showed immense difficulties in triggering field evaporation from a thick layer of ice was attributed to a large voltage drop in past attempts (Stephenson et al., 2018). Here, with thinner ice layers seem to allow for the generation of electrostatic fields that are sufficient to trigger field evaporation, including on the unsharpened specimens (approx. 100 µm in diameter) under high-intensity laser illumination. More experimental data acquired under controlled conditions and with systematic variations of the parameters will likely be necessary. Of particular interest is the strength of the electrostatic field at the specimen's apex and within the bulk of the specimen, and their influence on the field evaporated molecular species.

4.4.2   Analytical performance

In general, optimising the performance and yield of APT is about balancing the electric field and the temperature (Wada, 1984; Baptiste Gault et al., 2010; B. Gault et al., 2010; Yao et al., 2011;

Mancini et al., 2014). Even for more conventional materials than ice, the analytical limits (spatial resolution, sensitivity) of APT are typically not readily available or quantified (Lefebvre-Ulrikson et al., 2016). The quantification of carbon is a good example that has been treated multiple times and that is dependent both on the instrument performance and on the complexities of the field evaporation and dissociative behaviour of some of the emitted molecular ions (Sha et al., 1992; Thuvander et al., 2011; Peng et al., 2018, 2019). Similar studies of the dissociative behaviour of water-based molecular ions will be required, and these will need to combine experiments and atomistic simulations (David Zanuttini et al., 2017; Schwarz et al., 2020; Gault et al., 2016).

In addition, the possible penetration of the electrostatic field below the ice's surface could drive ion migration inside the specimen or towards its surface. The influence of the electrostatic field and associated gradients at the specimen's surface (Wang & Tsong, 1982) or in the bulk of the specimen (Greiwe et al., 2014) have been shown previously. The heat generated by the laser pulses can affect the specimen's outer shape (Figure 15) but also its interior due to possible diffusion that is driven by the inner electrostatic field. Since the shape of the emitter affects the ion projection (Hyde et al., 1994; Loi et al., 2013), changes in the specimen's shape over the course of the analysis would hence introduce uncertainties in the spatial reconstruction. In addition, the reconstruction protocols typically used across the community (Larson, 2013; Gault et al., 2011; Bas et al., 1995) make use of the atomic volume to calculate the depth increment. What volume should be used for various molecular ions? Little has been done in the community with respect to such a major issue, and only once has the data been reconstructed (Breen et al., 2013).

### 4.4.3 Effect of base temperature and of laser pulsing

To balance the electric field, one can increase the temperature at which the field evaporation takes place. At higher temperatures, the yield improves at the expense, typically, of the spatial resolution

(Müller, Gault, et al., 2011). This is how laser pulsing works, but this requires absorption of photons from the laser pulse that heat up the specimen, towards its apex. The optical absorption by metallic atom probe specimens has been studied in details (Robins et al., 1986; Sha et al., 2008; Vurpillot et al., 2009), including as a function of the wavelength (Houard et al., 2011, 2009). For ice, especially under high electric field, such studies are so far unchartered territories. What could be the effect of the wavelength or pulse duration on the field evaporation behaviour? Is the evaporation caused by thermal excitation or by electronic excitation after the absorption of a photon? Once again, these questions remain unanswered for the time being.

4.4.4 Ice's crystal structure

An additional level of complexity is that, at this stage, we do not know the crystallography of the ice that was generated – is it amorphous or crystalline? And if it is crystalline, which one of the many phases (Engel et al., 2018) are dominant? Does the very large electrostatic field generated at the apex cause a change in the structure itself? Indeed, APT specimens must sustain a high amount of stress during the analysis, because of the electrostatic pressure associated to the intense electric field (Rendulic & Müller, 1967). This stress can cause metallic specimens mechanically to fail during the analysis (Wilkes et al., 1972; Moy et al., 2011), and in some cases the low yield seem to make meaningful analysis simply impossible. Is this phenomenon relevant at all to frozen liquids? Under high pressures, and even at cryogenic temperatures, ice can deform (Wu & Prakash, 2015), which could lead to a change in the specimen's shape and potentially in the distribution of solutes.

There will be a need to define protocols allowing for confirmation of the amorphous or crystalline nature of the ice. Traditionally, assessing the structure may have been done by field ion microscopy for instance, but the scattered results in the literature do not give strong indications that this will be a simple endeavour (Stintz & Panitz, 1991, 1992; Stuve, 2012). Correlative workflows with TEM have been developed over the past decades (Loberg & Norden, 1968; Felfer et al., 2012; Herbig et al., 2015; Herbig, 2018; Liebscher et al., 2018; Makineni et al., 2018) as well as with transmission Kikuchi diffraction (Babinsky et al., 2014; Breen et al., 2017; Schwarz et al., 2018). Correlative protocols will now need to be adapted for cryo-transmission–Kikuchi diffraction or cryo-TEM to confirm the ice's crystal structure.

# 5 Conclusions

All new microscopy developments are underpinned by the ability to prepare specimens, and this study is hence a necessary step towards opening a new playing field for near-atomic scale analysis of solute effects in confined freezing, nano-objects, and molecular or biological materials in their native environment. Much remains to be done to grasp the physics of the field evaporation of frozen liquids, but the preliminary data obtained on two separate instruments from our two separate groups, in parallel, demonstrate the general feasibility of these analyses. Alternative approaches for the preparation of specimens suitable for both TEM and APT are currently under development and will surely keep blossoming in the coming years.

# 6 Acknowledgements

We thank Uwe Tezins, Christian Broß and Andreas Sturm for their support to the FIB and APT facilities at MPIE. Professor Roger Newman from the University of Toronto is acknowledged for providing the AgAu alloys. We are grateful for the financial support from the BMBF via the project UGSLIT and the Max-Planck Gesellschaft via the Laplace project. A.A.E., S.-H.K., L.T.S. and B.G. acknowledge financial support from the ERC-CoG-SHINE-771602. The funding by the

German Research Foundation (Deutsche Forschungsgemeinschaft, DFG) within the collaborative research center (CRC) 1333 (project number 358283783) is gratefully acknowledged.